\documentclass[%
 reprint,
 amsmath,amssymb,
 aps,
 prc
]{revtex4-1}

\usepackage[utf8]{inputenc} 
\usepackage[colorlinks=true,allcolors=blue]{hyperref} 
\usepackage[toc,page]{appendix} 
\usepackage[shortlabels]{enumitem} 
\usepackage{xspace}
\usepackage{xcolor}
\usepackage{braket}

\usepackage{graphicx} 
\usepackage{tabularx} 
\usepackage{dcolumn}  
\newcolumntype{d}[1]{D{.}{.}{#1}} 
\newcolumntype{R}{>{\raggedleft\arraybackslash}X} 

\usepackage{bm} 
\usepackage{mathtools} 

\begin{document}

\title{Effects of Quasiparticle-Vibration Coupling on Gamow-Teller Strength and
$\beta$ Decay with the Skyrme Proton-Neutron Finite-Amplitude Method}

\author{Qunqun Liu}
\email[]{qunqun@unc.edu}
\author{Jonathan Engel}
\email[]{engelj@physics.unc.edu}
\affiliation{Department of Physics and Astronomy, CB 3255, University of North
Carolina, Chapel Hill, North Carolina 27599-3255, USA\looseness=-1}
\author{Nobuo Hinohara}
\email{hinohara@nucl.ph.tsukuba.ac.jp}
\affiliation{Center for Computational Sciences, University of Tsukuba, Tsukuba
305-8577, Japan} 
\affiliation{Faculty of Pure and Applied Sciences, University of Tsukuba,
Tsukuba 305-8571, Japan}
\affiliation{Facility for Rare Isotope Beams, Michigan State University, East
Lansing, Michigan 48824, USA}
\author{Markus Kortelainen}
\email{markus.kortelainen@jyu.fi}
\affiliation{Department of Physics, University of Jyv\"askyl\"a, P.O. Box 35 
(YFL), FI-40014 Jyv\"askyl\"a, Finland}

\date{\today}

\begin{abstract}
We adapt the proton-neutron finite-amplitude method, which in its original form
is an efficient implementation of the Skyrme quasiparticle random phase
approximation, to include the coupling of quasiparticles to like-particle
phonons.  The approach allows us to add beyond-QRPA correlations to computations
of Gamow-Teller strength and $\beta$-decay rates in deformed nuclei for the
first time.  We test the approach in several deformed isotopes for which
measured strength distributions are available.  The additional correlations
dramatically improve agreement with the data, and will lead to improved global
$\beta$-decay rates. 
\end{abstract}

\maketitle

\section{Introduction}
The \textit{r} process, which is responsible for synthesizing many of the heavy
elements, is not fully understood \cite{Kajino2019}.  It is thought at present
to occur primarily in neutron-star mergers, but may also take place in
supernovae.  To pin down the conditions under which rapid neutron capture and
$\beta$ decay can lead to observed isotopic abundances, we need to understand
the properties of nuclei that are too neutron-rich to be made in laboratories.
Among the most important properties are $\beta$-decay rates.

Computing these rates in all neutron-rich isotopes is a difficult undertaking.
Though \textit{ab initio} methods for solving the nuclear many-body problem have
made great strides \cite{Hergert2020,hagen14,Navratil2016,Lynn2019}, they have
not yet been extended to heavy nuclei far from closed shells.  The best approach
for now is energy-density-functional theory, in particular its version of linear
response, which relates density oscillations to transition rates.  References
\cite{Mustonen2016,Shafer2016,Ney20,Marketin2015} have applied the
charge-changing version of the Skyrme or relativistic quasiparticle random-phase
approximation (QRPA) to produce tables of $\beta$-decay rates in thousands of
isotopes.  The method can be used either to compute transition rates directly
through the diagonalization of a QRPA Hamiltonian matrix, or to extract the
rates from response functions.  The Hamiltonian matrix is so time-consuming to
build, however, that the matrix approach in Ref.\ \cite{Marketin2015} required
the assumption of spherical symmetry to obtain the thousands of rates
needed for simulating nucleosynthesis.  

With the advent of the finite amplitude method (FAM)
\cite{Nakatsukasa07,avogadro11} for computing QRPA linear response, calculations
of strength distributions in deformed isotopes
\cite{Stoitsov11a,Kortelainen2015,Hinohara13} became straightforward.  The
global tables in Refs.\ \cite{Mustonen2016,Ney20} were obtained with the
charge-changing version of this approach, called the proton-neutron FAM (pnFAM)
\cite{Mustonen2014a}, and the assumption only of axial symmetry.  The QRPA
within density-functional theory has limitations, however, no matter how it is
formulated.  A linear response produced by oscillations of a mean-field is at
best an adiabatic approximation, correct only at an oscillation frequency of
zero, even if the time-independent functional on which the response is based is
exact (which it never is).  One can obtain a more realistic frequency-dependent
response by coupling the quantized oscillations --- phonons --- to the
quasiparticles that compose the phonons, or to other phonons.  The first
option, when developed systematically, leads in lowest order to the
time-blocking approximation
\cite{Kamerdzhiev2006,Litvinova07,lit08a,Tselyaev2016}, which is equivalent to
an density-functional version of what is called the
``quasiparticle-vibration-coupling'' model \cite{Colo1994,Li2022}.  Within
Skyrme density-functional theory, this approximation has been used in a limited
number of spherical nuclei to compute Gamow-Teller strength distributions
\cite{Niu2016} and $\beta$-decay rates \cite{Niu2018}.  The phonons are
like-particle excitations that are emitted and reabsorbed by the proton and
neutron quasiparticles that underlie the excitations of the charge-changing
QRPA.  To make the picture produce the correct zero-frequency response, which is
determined completely by the static Skyrme functional, one can employ the
subtraction procedure first proposed in Ref.\ \cite{Tselyaev2013}.

The coupling of quasiparticles to phonons significantly improves agreement
with data in spherical nuclei.  One would like to use the method in global
calculations of $\beta$-decay rates but faces the same problem encountered by
the QRPA itself a few years ago: the usual implementation is through a
Hamiltonian matrix, which is too time-consuming to construct when spherical
symmetry cannot be exploited.  Because the vast majority of nuclei are deformed,
we thus need a different formalism.  An extension of the pnFAM is the obvious
choice.  A formalism for the extension of the like-particle FAM with
relativistic density functionals was developed, though not applied, in Refs.\
\cite{Zhang2022,Litvinova2022}.  Here we both show how to extend the pnFAM to
include the coupling of quasiparticles to like-particle phonons and use the
extension to compute Gamow-Teller distributions in several deformed isotopes,
finding the agreement with experiment to be dramatically better than in the
original pnFAM.

Treating $\beta$-decay in our new approach is a separate task because its rates
are sensitive to small amounts of low-lying Gamow-Teller strength, which the
Skyrme functionals that we have used in the past were adjusted to reproduce
within the ordinary pnFAM.  The parameters of the functionals must thus be refit
before an improved table of rates can be created.  We can, of course,
demonstrate the effects of coupling quasiparticles to phonons on a few
representative $\beta$-decay rates without refitting, and we do that here.  The
global application of the approach faces an additional obstacle, however: to
obtain the interaction between quasiparticles and phonons, we apply the
like-particle FAM in a way that will be hard to automate for use in the
thousands of isotopes for which we need $\beta$-decay rates.  Towards the end of
this paper, we discuss relatively straightforward steps that will remedy the
problem, deferring their implementation and the Skyrme-parameter refitting to
the future.

The rest of the paper is structured as follows:  Section II presents our method
for adding the coupling of quasiparticles and phonons to the pnFAM and discusses
subtleties that arise in deformed nuclei.  Section III presents Gamow-Teller
distributions in $^{76}$Ge, $^{82}$Se, and $^{150}$Nd, and compares them with
experimental data from charge-exchange reactions.  Section IV presents
$\beta$-decay rates in 12 deformed isotopes, showing that the new physics
usually increases those rates.  Section V contains a roadmap of sorts for the
computation of $\beta$-decay rates in all unstable nuclei, a discussion of the
explicit treatment of correlations within a density-functional framework, and a
conclusion.

\section{Formalism}

Our goal is to compute the strength distribution produced by a charge-changing
operator $F$:
\begin{equation}
\label{eq:Fform}
F = \sum_{pn} f_{pn} a^\dag_p a_{n} \,,
\end{equation}
where the coefficients $f_{pn}$ are arbitrary.  The strength distribution can be
written in the form
\begin{equation}
\label{eq:strdef}
\frac{d B(F,\omega)}{d \omega}=-\frac{1}{\pi} \operatorname{Im} S_F(\omega) \,, 
\end{equation}
where $\omega$ is the frequency with which $F$ perturbs the nucleus (or
equivalently, the amount of energy it supplies) and the ``response function''
$S_F(\omega)$ is
\begin{equation}
\label{eq:respdef}
S_F(\omega) =-\sum_M\left(\frac{|\langle M|F| 0\rangle|^2}
{\Omega_M-\omega}-\frac{|\langle M|F^{\dag}|
0\rangle|^2}{\Omega_M+\omega}\right) \,. 
\end{equation}
Here $\ket{0}$ is the ground state of the initial nucleus and the $\ket{M} $'s
are states in the final nuclei populated by single-charge-exchange processes
with energies $\Omega_M$ above the energy of $\ket{0}$.   The perturbing
frequency/energy $\omega$ is a complex number a distance $\Delta$ above the real
axis.  Though $\Delta$ is supposed to be taken to zero at the end of any
calculation, non-zero values supply widths to peaks that mock up effects of the
continuum (which our calculations neglect).

After an HFB transformation, without proton-neutron mixing, the operator $F$ can
be written as
\begin{equation}
\label{eq:Fqp}
F = \sum_{\pi\nu} \left( F^{20}_{\pi \nu} \alpha^\dag_\pi \alpha^\dag_\nu +
F^{02}_{\pi \nu} \alpha_\nu \alpha_\pi  + F^{11}_{\pi \nu}
\alpha^\dag_\pi \alpha_\nu + F^{11}_{\nu \pi} \alpha^\dag_\nu \alpha_\pi \right)\,, 
\end{equation}
where the Greek letters $\pi$ and $\nu$ label proton and neutron quasiparticle
orbitals, and $\alpha^\dag$ and $\alpha$ represent operators that create and
annihilate quasiparticles.  The $F^{ij}_{\pi \nu}$ in Eq.\ (\ref{eq:Fqp}) depend
on the $f$'s in Eq.  \eqref{eq:Fform} and the HFB matrices $U$ and $V$
\cite{rin04} that specify the transformation from particles to quasiparticles.
In the pnFAM, the response function can be written in the form
\begin{equation}
\label{eq:FAMstr}
\begin{aligned}
S_F(\omega) & =\sum_{\pi \nu}\left[F_{\pi \nu}^{20 *} X_{\pi \nu}(\omega)
+F_{\pi \nu}^{02 *} Y_{\pi \nu}(\omega)\right] \,, 
\end{aligned}
\end{equation}
where the $X$'s and $Y$'s are the fluctuation amplitudes in density-matrix
elements induced by the action of $F$, applied at frequency $\omega$.  The pnFAM
equations for these amplitudes are \cite{Mustonen2014a}
\begin{equation}
\label{eq:pnFAMeqs}
\begin{aligned}
 X_{\pi \nu}(\omega) &=-\frac{\delta H_{\pi
\nu}^{20}(\omega)+F_{\pi \nu}^{20}(\omega)}{\varepsilon_\pi+\varepsilon_v-\omega} \\
 Y_{\pi \nu}(\omega) &=-\frac{\delta H_{\pi
 \nu}^{02}(\omega)+F_{\pi \nu}^{02}(\omega)}{\varepsilon_\pi+\varepsilon_v+\omega} \,. 
\end{aligned}
\end{equation}
Here the $\varepsilon$'s are quasiparticle energies and $\delta H_{\pi
\nu}^{20}$ and $\delta H_{\pi \nu}^{02}$ are the pieces of the fluctuating
generalized HFB mean-field Hamiltonian that multiply pairs of quasiparticle
creation and annihilation operators in the same way as do $F^{20}_{\pi\nu} $ and
$F^{02}_{\pi\nu}$ in Eq.\ \eqref{eq:Fqp}.  Because $\delta H^{20}$ and $\delta
H^{02}$ depend on the $X$'s and $Y$'s, Eqs.\ \eqref{eq:pnFAMeqs} are most easily
solved through iteration.

In the QRPA, the states $\ket{M}$ are simple two-quasiparticle and two-quasihole
excitations of the ground state $\ket{0}$.  We want to make these simple states
more realistic by allowing them to mix with states that include coherent
like-particle excitations.   The most straightforward way of doing that is to
allow the emission and re-absorption of like-particle QRPA phonons by one of the
quasiparticles in the two-quasiparticle excitation of $\ket{0}$, or the exchange
of such a phonon between the two.  When only one phonon is allowed to exist at a
time within a time-ordered picture for the two-quasiparticle propagator, we end
up with the quasiparticle-vibration coupling model \cite{Niu2016} or,
equivalently, the time-blocking approximation \cite{Litvinova07,Tselyaev2016}. 

The modifications to the FAM equations induced by the quasiparticle-phonon
coupling can be derived in a number of ways.  One can, for example, follow the
equations of motion method for charge-changing excitations, for example, using
the ansatz
\begin{equation}
\label{eq:eqmotion}
\ket{M} = Q_M^\dag  \ket{0} \,,
\end{equation}
with
\begin{equation}
\label{eq:exc-op}
\begin{aligned}
Q^\dag_M & = \sum_{\pi \nu} \big( X^M_{\pi \nu} \alpha^\dag_\pi
\alpha^\dag_\nu - Y^M_{\pi \nu} \alpha_\nu \alpha_\pi  \\
& \qquad  + \sum_{N} \tilde{X}^M_{\pi \nu N}
\alpha^\dag_\pi \alpha^\dag_\nu \mathcal{Q}^\dag_N - \tilde{Y}^M_{\pi \nu N} 
\mathcal{Q}_N \alpha_\nu \alpha_\pi \big) \,.  \\[1cm]
\end{aligned}
\end{equation}
Here $\mathcal{Q}^\dag_N$ creates the $N^{\text{th}}$ like-particle phonon (with
non-negative energy) in the usual like-particle QRPA, the $X$'s and $Y$'s are
now charge-changing QRPA-level amplitudes, and the $\tilde{X}$'s and
$\tilde{Y}$'s are beyond-QRPA amplitudes that specify the ways in which
quasiparticles couple to phonons.  One might also include terms in which pairs
of like quasiparticles couple to charge-changing phonons, but those are less
important \cite{Robin18} and we neglect them here.

\begin{figure}[t]
\centering
\includegraphics[width=.74\columnwidth]{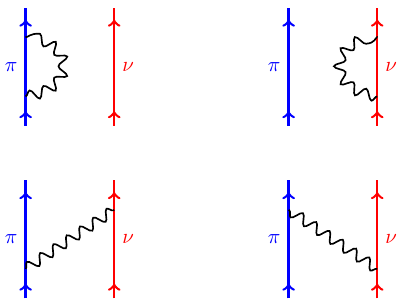}
\caption{Diagrams illustrating the four contributions to the ``spreading
matrix'' $W$ in Eq.\ \eqref{eq:Wform}.  Blue lines represent protons and red
lines neutrons. Black squiggly lines represent phonons, either emitted and
re-absorbed by a single nucleon or exchanged between nucleons.}
\label{fig:diagrams}
\end{figure}

When the complicated amplitudes $\tilde{X}$ and $\tilde{Y}$ are eliminated in
the usual way in favor of a propagator in the space of complicated excitations,
one finds, instead of Eq.\ \eqref{eq:pnFAMeqs}, relations that we refer to as
the pnFAM* equations:
\begin{equation}
\label{eq:modpnFAMeqs}
\begin{aligned}
X_{\pi \nu}(\omega) &=-\frac{\delta H_{\pi \nu}^{20}+\left[\tilde{W}(\omega)
X(\omega)\right]_{\pi
\nu}+F_{\pi \nu}^{20}}{\varepsilon_\pi +\varepsilon_\nu-\omega} \\
Y_{\pi \nu}(\omega) &=-\frac{\delta H_{\pi \nu}^{02}+\left[\tilde{W}^*(-\omega)
Y(\omega)\right]_{\pi \nu}+F_{\pi \nu}^{02}}{\varepsilon_\pi +\varepsilon_\nu+\omega} \,, 
\end{aligned}
\end{equation}
where 
\begin{equation}
\label{eq:Wtilde}
\tilde{W}(\omega) = W(\omega) - W(0) \,,
\end{equation}
and
\begin{equation}
\label{eq:longhand}
\begin{aligned}
\left[\tilde{W}(\omega) X(\omega)\right]_{\pi \nu} &=\sum_{\pi'\nu'}
\tilde{W}_{\pi \nu, \pi ' \nu'}(\omega)
X_{\pi' \nu'}(\omega) \\
\left[\tilde{W}^*(-\omega) Y(\omega)\right]_{\pi \nu} &=\sum_{\pi'\nu'} 
\tilde{W}_{\pi \nu, \pi' \nu'}^*(-\omega) 
Y_{\pi' \nu'}(\omega) \,, 
\end{aligned}
\end{equation}
The matrix $W$, which includes a phonon-loop correction to single-quasiparticle
energies and a phonon-exchange interaction (see Fig.\ \ref{fig:diagrams}) is
given by
\begin{widetext}
   \begin{equation}
   \label{eq:Wform}
   \begin{aligned}
   W&_{\pi \nu, \pi' \nu'}(\omega) =\sum_N \Bigg\{
   \sum_{\pi_1}\left\langle \pi |H| \pi_1, N\right\rangle
   \frac{1}{\omega-\left[\omega_N+\left(\varepsilon_{\pi_1}+\varepsilon_\nu \right)\right]} 
  \left\langle \pi'|H|\pi_1, N\right\rangle^*\delta_{\nu' \nu}\\
   & \qquad \qquad \qquad \qquad \qquad 
   +\sum_{\nu_1}\left\langle \nu|H| \nu_1, N\right\rangle
   \frac{1}{\omega-\left[\omega_N+\left(\varepsilon_\pi+\varepsilon_{\nu_1}\right)\right]}\left\langle
   \nu'|H| \nu_1, N\right\rangle^* \delta_{\pi' \pi} \\
   & \qquad \qquad +\left\langle \pi|H| \pi', N\right\rangle
   \frac{1}{\omega-\left[\omega_N+\left(\varepsilon_{\pi'}+
   \varepsilon_\nu\right)\right]}\left\langle \nu'|H| \nu,
   N\right\rangle^* +\left\langle \nu|H| \nu', N\right\rangle
   \frac{1}{\omega-\left[\omega_N+\left(\varepsilon_\pi+
   \varepsilon_{\nu'}\right)\right]}
   \left\langle \pi'|H| \pi, N\right\rangle^*\Bigg\} \,. 
   \end{aligned}
   \end{equation}
\end{widetext}
Here $\omega_N$ is the energy of the $N^{\text{th}}$ like particle phonon.  Each
of the four terms in Eq.\ \eqref{eq:Wform} corresponds to one of the diagrams in
Fig.\ \ref{fig:diagrams}.  Using $\tilde{W}(\omega)$ instead of $W(\omega)$ in
Eq.\ \eqref{eq:modpnFAMeqs} implements the subtraction procedure
\cite{Tselyaev2013} that guarantees that the static response is the same as in
the unmodified pnFAM \cite{gam15}.

Within the expression in Eq.\ \eqref{eq:Wform} for $W$, which is sometimes
called the ``spreading matrix,'' are quasiparticle-phonon vertices, matrix
elements of the Hamiltonian with a single quasiparticle on one side and another
quasiparticle of the same type plus a phonon on the other.  The phonons can be
excited by a like-particle one-body operator $G$:
\begin{equation}
\label{eq:like-op}
G = \sum_{pp'} G_{pp'} a^\dag_p a_{p'} + \sum_{nn'} G_{nn'} a^\dag_n a_{n'}
\,, 
\end{equation}
where $p,p'$ label proton orbitals and $n,n'$ label neutron orbitals.  In
like-particle linear-response theory, $G$ generates fluctuations $\delta H$ in
the like-particle HFB Hamiltonian.  The quasiparticle-phonon vertices can be
related to $\delta H^{11}$, the coefficients of operators of the form
$\alpha^\dag_\nu \alpha_{\nu'}$ and $\alpha^\dag_\pi \alpha_{\pi'}$ in the fluctuating
like-particle HFB Hamiltonian, through a contour integral \cite{Hinohara13} or,
as Ref.\ \cite{Zhang2022} shows, through the relation 
\begin{equation}
\label{eq:vertex}
\braket{\beta|H|\beta_1,N} = i \lim_{\Delta \to 0}
\frac{\Delta}{\braket{N|G|0}} 
\delta H^{11}_{\beta\beta_1}(\Omega_N + i \Delta)  \,.  
\end{equation}
Here $\beta$ and $\beta_1$ can represent either proton or neutron quasiparticle
orbitals as long as they are of the same type, and $\braket{N|G|0}$ is the
matrix element connecting the ground state with the $N^\text{th}$ phonon.  In
the like-particle FAM, this last quantity can be extracted up to an arbitrary
and irrelevant phase from the like-particle response function $S_G$, 
\begin{equation}
\label{eq:respdef-like}
S_G(\omega) =-\sum_N\left(\frac{|\langle N|G| 0\rangle|^2}
{\omega_N-\omega}-\frac{|\langle N|G^\dag |
0\rangle|^2}{\omega_N+\omega}\right) \,, 
\end{equation}
as
\begin{equation}
\label{eq:GN0}
\braket{N|G|0} = \lim_{\Delta \to 0} \sqrt{i \Delta S_G(\omega_N + i \Delta)}
\,. 
\end{equation}

In this paper, in place of the generic charge-changing operator $F$ we will
usually use the components of the Gamow-Teller operator,
\begin{equation}
\label{eq:Fhere}
F_K = \sum_{i} \sigma_K(i) \tau_-(i) \,,
\end{equation}
where the sum is over nucleons, $K$ labels the projection of the angular
momentum along the $z$ axis, and $\tau_- \equiv \frac{1}{2} (\tau_x - i \tau_y)$
--- note the non-standard normalization --- turns neutron states into proton
states while annihilating the latter.   In place of $G$ we will use the
multipole operators
\begin{equation}
\label{eq:Ghere}
\begin{aligned}
G^{T=0}_{LK} & = \sum_{i} r_i^L Y_{LK}(\theta_i,\varphi_i)\\
G^{T=1}_{LK} & = \sum_{i} r_i^L Y_{LK} (\theta_i,\varphi_i) \tau_z (i) \,, 
\end{aligned}
\end{equation}
where $\tau_z$ is 1 for neutrons and $-1$ for protons.  Here $L$ is any integer,
$r$ is the radial coordinate, and $Y_{LK}$ is the spherical harmonic with angular
momentum $L$ and $z$-projection $K$. For a given charge-changing projection $K$,
like-particle phonons created by operators $G^T_{L,K'}$ with any value of $K'$
can play a role.  To eliminate the spurious states corresponding to translation
and rotation, we simply discard phonons with an energy less than 0.5 MeV.  

\section{Gamow-Teller Strength Functions}

The computation of strength functions proceeds as follows:  First, we use the
code HFBTHO \cite{Stoitsov2005,Stoitsov20131592} to solve the HFB equations for
arbitrary Skyrme functionals.  In the work here, we use a single-particle space
consisting of 16 shells.  With quasiparticle energies and wave functions in
hand, we then run the like-particle FAM enough times to generate all states
excited by the natural-parity operators in Eq.\ \eqref{eq:Ghere} with $L \leq
6$.  (our code is not yet set up for the less-important unnatural-parity
excitations).  To isolate those states, in each multipole (specified by $L$ and
$K$ in spherical nuclei, and parity and $K$ in deformed nuclei) we solve the FAM
equations for 40 values of the excitation frequency $\omega$ between 0 and 20
MeV, with an imaginary component of 0.5 MeV, to identify the peaks in the
strength function.  For each such peak we run the FAM several times again, with
the real part of $\omega$ at the energy of the peak and the imaginary part
decreasing towards zero, and use Eq.\ \eqref{eq:vertex} to compute the
quasiparticle-phonon vertices. 

The final step is to use Eq.\ \eqref{eq:Wform} to construct the spreading matrix
$W$ and solve the pnFAM* equations in Eq.\ \eqref{eq:modpnFAMeqs}.  Rather than
store the four-index matrix $W$, we store the simpler quasiparticle-phonon
vertices and construct $W$ on the fly as needed.  After obtaining $X_{\pi \nu}
(\omega)$ and $Y_{\pi \nu} (\omega)$ for many values of $\omega$ near the real
axis, we use Eqs.\ \eqref{eq:strdef} and \eqref{eq:FAMstr} to construct first
the response and then the Gamow-Teller strength distribution.

In what follows, we display results for the strength distribution obtained with
the Skyrme functional SGII \cite{vangiai1981,vangiai1981a}, which was designed
for spin-isospin excitations, with the same effective interaction in the
time-even and time-odd channels.  The pairing is mixed surface-volume, with
strengths of $-265.25$ MeV \!fm$^3$ for neutrons and $-340.0625$ MeV \!fm$^3$ for
protons, and a pairing window cut off at 60 MeV. Distributions have been
measured in several nuclei that can undergo $\beta \beta$ decay (to provide
information that bears on the matrix elements that govern the two-neutrino and
neutrinoless versions of that process), and we begin with the $^{82}$Se.  Our
HFBTHO calculations predict the nucleus to have mild axially-symmetric
deformation, with the quadrupole-deformation parameter,
\begin{equation}
\label{eq:betadef}
\beta = \sqrt{\frac{\pi}{5}} \frac{Q_2}{R^2} \,,
\end{equation}
given by $\beta = 0.13$.  Here $Q_2$ is the usual quadrupole moment and $R^2$
the mean-square nuclear radius.  

\begin{figure}[t]
\centering
\includegraphics[width=\columnwidth]{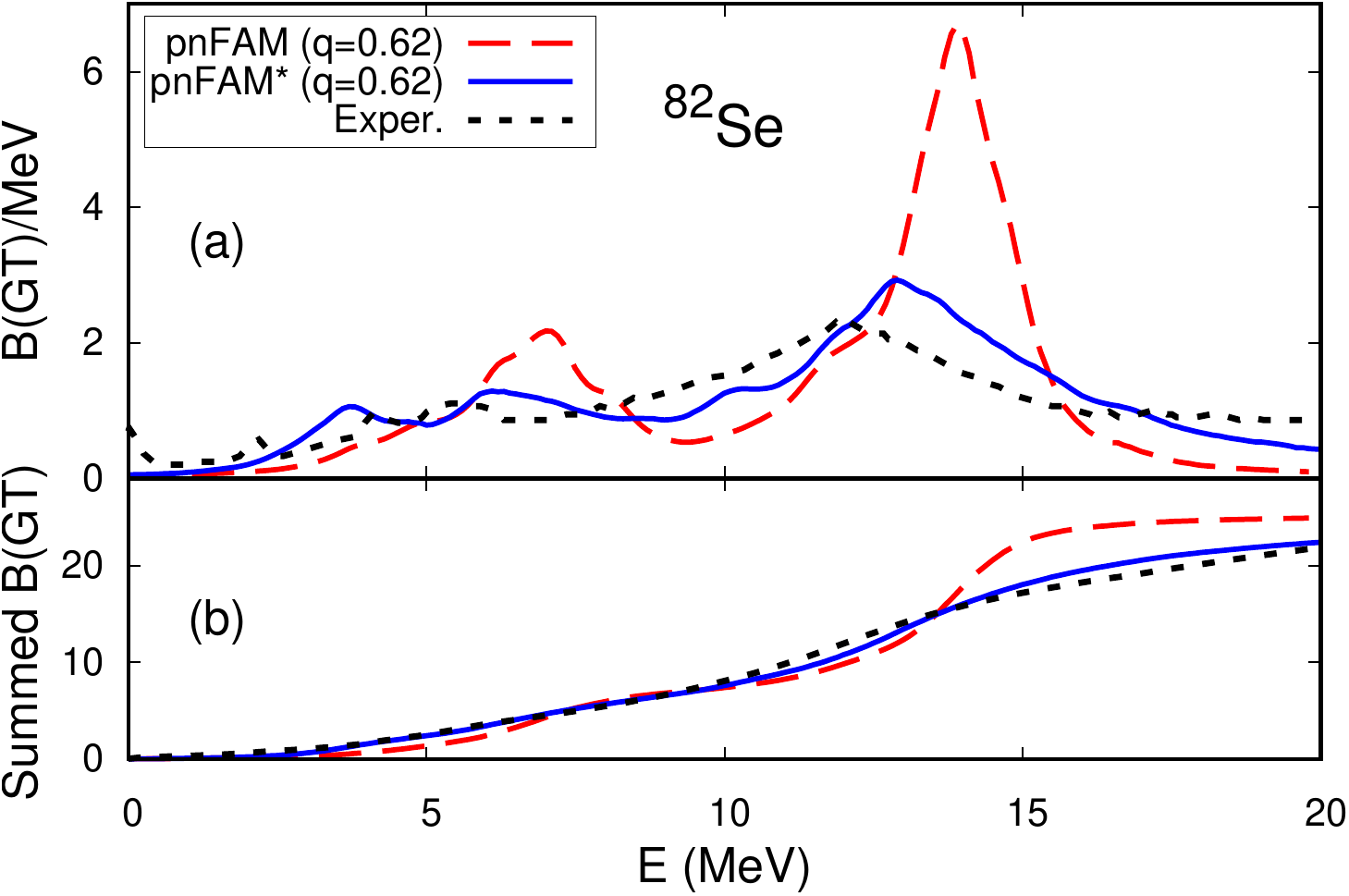}
\caption{(a) Gamow-Teller strength distributions for the nucleus $^{82}$Se.  The
red dashed and solid blue curves, representing results calculated with the pnFAM
and pnFAM* (see text), have been scaled by 0.62.  The black curve, from Ref.\
\cite{Madey89} is experimental data.  In obtaining each curve, we chose the
imaginary part $\Delta$ of the perturbing frequency to be 0.5 MeV. (b)
Integrated Gamow-Teller strength.}
\label{fig:se-qvc}
\end{figure}

Figure \ref{fig:se-qvc} shows the distributions both of the strength itself and
the summed strength, with each scaled by a factor of 0.62 that would correspond
to an effective value for the axial-vector coupling constant $g_A$ of $1.0$ in
$\beta$ decay calculations.  The figure also displays the measured strength
\cite{Madey89}, the overall normalization of which is highly uncertain.  The
most salient feature of the pnFAM* strength is that it is spread, and thus
agrees much better with the experimental distribution.  It is not very quenched,
however, which is why we scale it; evidently the absence of multi-phonon
emission and reabsorption in $W$ is responsible for the small quenching.  (As in
most extended QRPAs, the Ikeda sum rule is preserved, so that any  quenching of
low-energy strength implies a long tail at high energies, as in Ref.\
\cite{Gambacurta2020}.) Both the spreading and lack of significant quenching
with the coupling of quasiparticles to vibrations are consistent with results in
spherical nuclei \cite{Niu2016,Gambacurta2020}.  But, in a promising sign for
the ability of these calculations to improve on pnFAM calculations of
$\beta$-decay rates, the figure shows the pnFAM* low-lying strength, by far the
most important for $\beta$ decay, to be much closer to that of experiment.

\begin{figure}[t]
\centering
\includegraphics[width=\columnwidth]{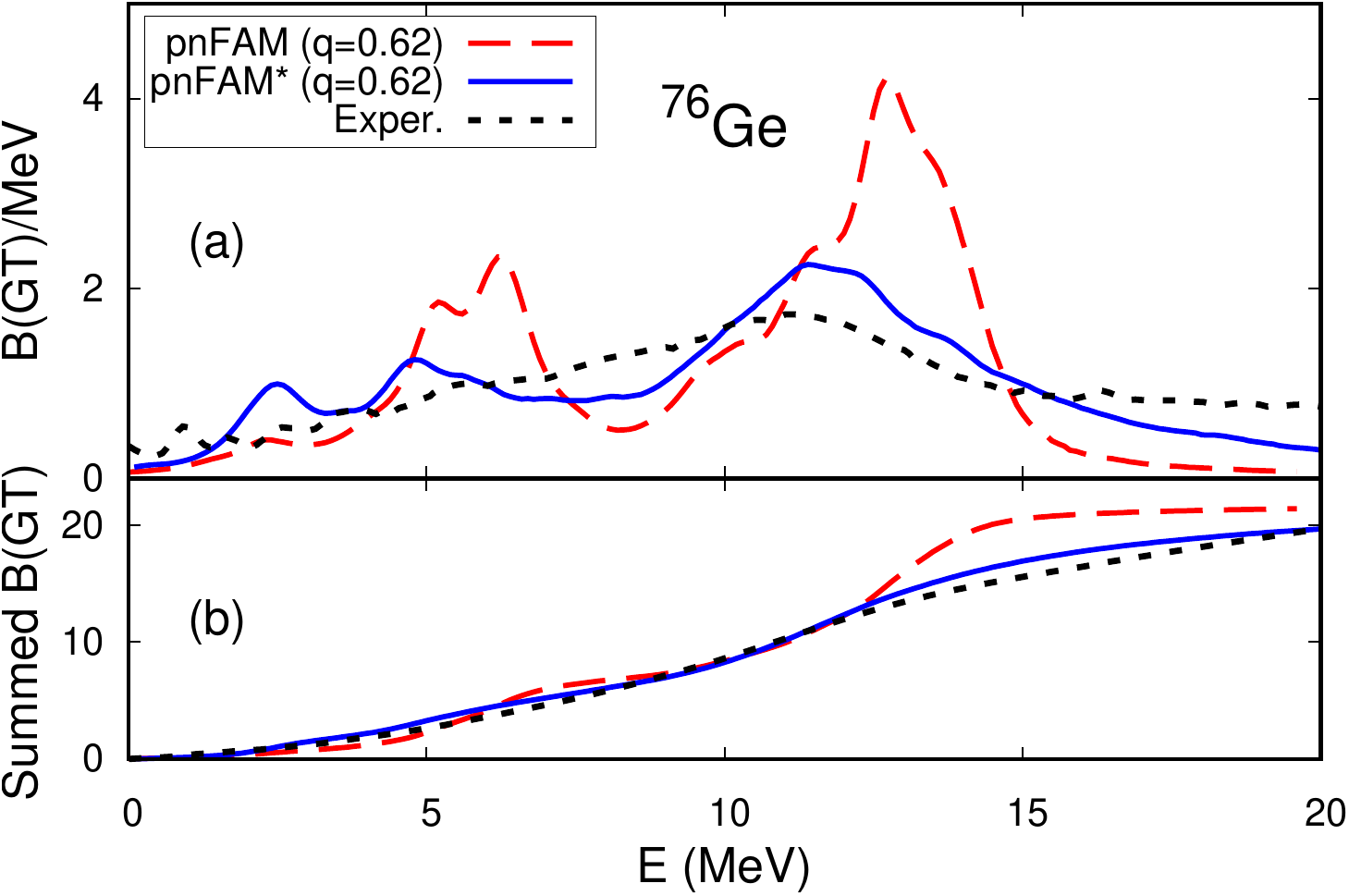}
\caption{Same as in Fig.\ \ref{fig:se-qvc} but for $^{76}$Ge.}
\label{fig:ge-qvc}
\end{figure}

\begin{figure}[b]
\centering
\includegraphics[width=\columnwidth]{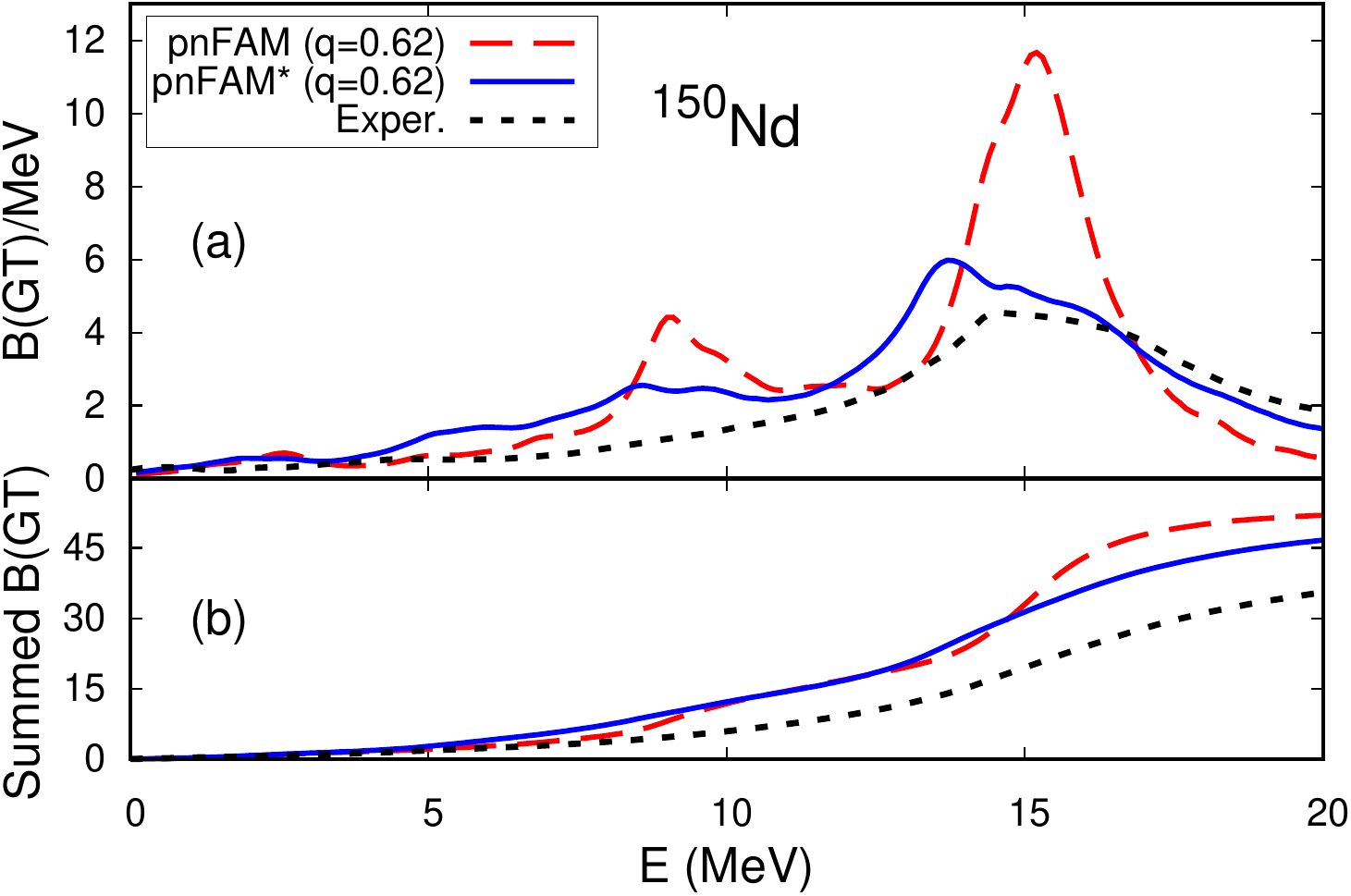}
\caption{Same as in Fig.\ \ref{fig:se-qvc} but for $^{150}$Nd, with the
experimental data from Ref.\ \cite{Guess2011}.}
\label{fig:nd-qvc}
\end{figure}

Figure \ref{fig:ge-qvc} shows the same curves as Fig.\ \ref{fig:se-qvc} but for
the nucleus $^{76}$Ge.  Recent work, both experimental
\cite{Toh13,Ayangeakaa2019} and theoretical \cite{Rodriguez2017}, suggests that
this nucleus is triaxial, but we (are forced to) treat it as axially symmetric,
with a deformation parameter from HFBTHO of $\beta = 0.12$.  The same features
are apparent here as in Fig.\ \ref{fig:se-qvc}.  

Figure \ref{fig:nd-qvc} shows the curves for the significantly heavier and more
deformed nucleus $^{150}$Nd, which in our calculations has $\beta = 0.21$.
Again, the pnFAM* improves agreement with experiment overall, but an even
stronger quenching than the 0.62 we apply is called for. 
  
The addition of the spreading matrix $W$ to the pnFAM slows the computation,
both because of the time required to evaluate $W$ and because the values for
$X_{\pi \nu}(\omega)$ and $Y_{\pi \nu}(\omega)$ in the iterative solution of
Eqs.\ \eqref{eq:modpnFAMeqs} converge more slowly than do those in Eqs.\
\eqref{eq:pnFAMeqs}.  To speed the computation, we would like to include as few
like-particle phonons as possible in the spreading matrix in Eq.\
\eqref{eq:Wform}.  But how do we decide which are the most important and how
many do we include?  These questions were addressed in spherical nuclei in Ref.\
\cite{Tselyaev2017}.  The authors showed that to evaluate the importance of the
$N^{\text{th}}$ phonon one can reliably use the ratio of the expectation value
$\braket{V}_N$ of the interaction in the phonon state to the phonon energy
$\omega_N$: 
\begin{equation}
\label{eq:vN}
v_N \equiv \frac{\braket{V}_N}{\omega_N} \,,
\end{equation}
This measure, justified carefully in Ref.\ \cite{Tselyaev2017}, reflects the
perturbative phonon exchange, which accentuates the importance of low-energy
phonons, and the relation between $\braket{V_N}$ and the quasiparticle-phonon
vertices in Eq.\ \eqref{eq:vertex}.   After some manipulation, $v_N$ can be
written in the form 
\begin{equation}
\label{eq:VN}
v_N = 1  - \frac{1}{\omega_N}  
\sum_{\alpha \beta} \left( \varepsilon_\alpha + \varepsilon_\beta \right) 
\left( \left|\mathcal{X}^N_{\alpha \beta}\right|^2 
+ \left| \mathcal{Y}^N_{\alpha \beta} \right|^2 \right) \,. 
\end{equation}
Here the $\mathcal{X}$'s and $\mathcal{Y}$'s are the like-particle QRPA analogs
of pnQRPA $X$'s and $Y$'s in Eq.\ \eqref{eq:exc-op}, and the indices $\alpha$
and $\beta$ run over all pairs in which both label protons or both label
neutrons.  Equations (24) in Ref.\ \cite{Hinohara13} implies that we can extract
the $\mathcal{X}$'s and $\mathcal{Y}$'s as the limits, 
\begin{equation}
\label{eq:extractxy}
\begin{aligned}
\mathcal{X}^N_{\alpha \beta} & = - i \lim_{\Delta \to 0}
\frac{\Delta}{\braket{N|G|0}} \mathcal{X}_{\alpha \beta} (\omega_N + i \Delta)\\ 
\mathcal{Y}^N_{\alpha \beta} & = - i \lim_{\Delta \to 0}
\frac{\Delta}{\braket{N|G|0}} \mathcal{Y}_{\alpha \beta} (\omega_N + i \Delta)
\,. 
\end{aligned}
\end{equation}
Here $G$ is one of the multipole operators in Eq.\ \eqref{eq:Ghere} and the
$\mathcal{X}_{\alpha \beta}(\omega)$ and $\mathcal{Y}_{\alpha \beta}(\omega)$
are the like-particle FAM amplitudes, i.e.\ the analogs of the pnFAM $X$'s and
$Y$'s in Eq.\ \eqref{eq:pnFAMeqs}. 

\begin{figure}[t]
\centering
\includegraphics[width=\columnwidth]{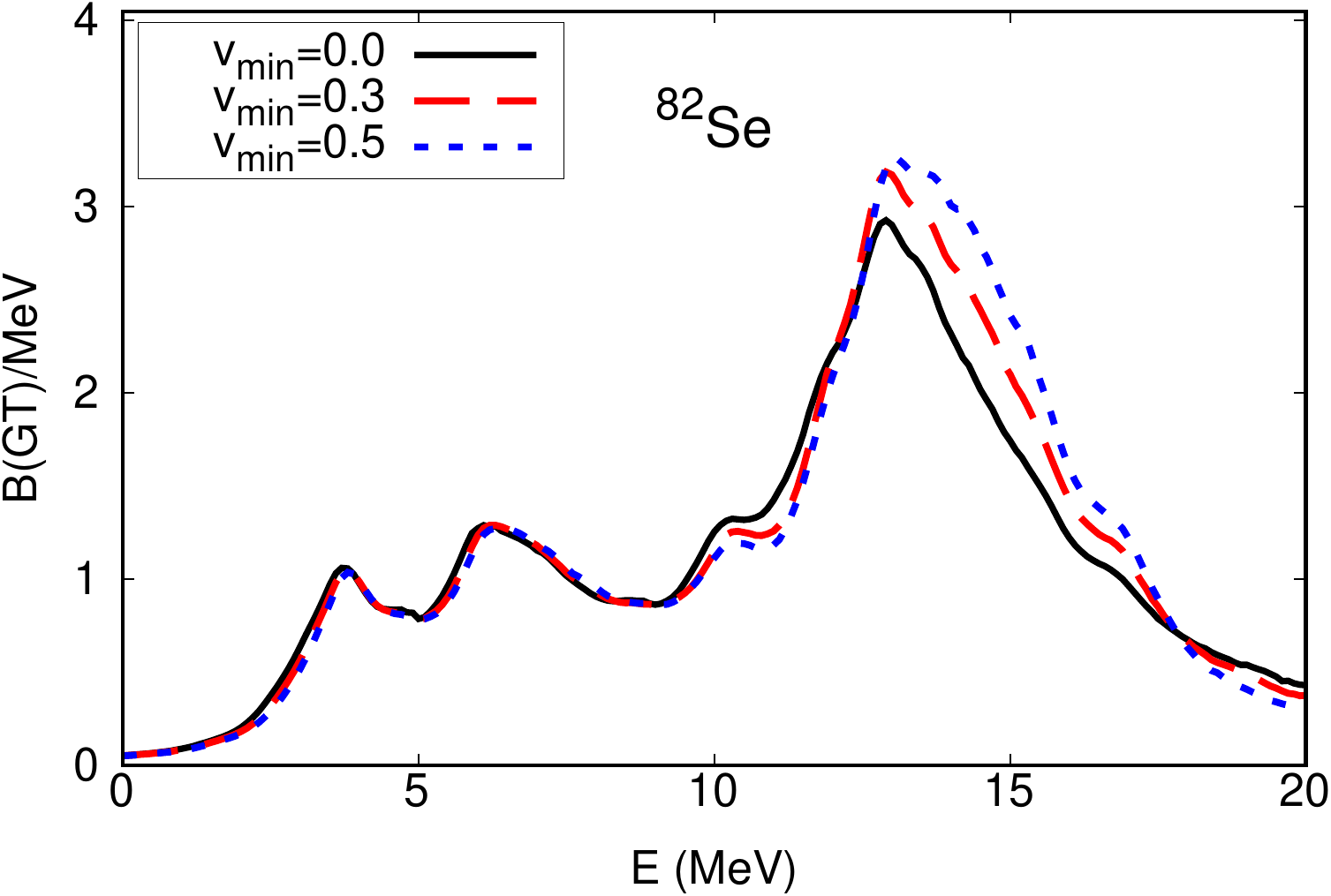}
\caption{Gamow-Teller strength distribution in $^{82}$Se obtained in the pnFAM*,
with several values for the cutoff parameter $v_{\text{min}}$.  The lower that
quantity, the more phonons the calculation includes.  As before, we include all
natural-parity multipoles with $L \leq 6$.}
\label{fig:se-cut}
\end{figure}

Figure \ref{fig:se-cut} shows the results in $^{82}$Se of requiring that $|v_N|$
be larger than a critical value $v_\text{min}$ as that quantity decreases.  At
the smallest value, $v_\text{min} = 0$, there is no truncation and the
calculation includes all the 150 phonons corresponding to distinct maxima in the
like-particle strength functions, when plotted with $\Delta = 0.5$ MeV.  At the
values $v_\text{min} = 0.3$, and $0.5$, the calculation includes 62 and 35
phonons, respectively.  Although truncation has a noticeable effect in the giant
Gamow-Teller resonance, even the most dramatic one does very little at low
energies.  That means that when calculating $\beta$ decay rates, we can expect
to get away with relatively few phonons.  The effects of truncation in $^{76}$Ge
and $^{150}$Nd are similar.  

\begin{figure}[t]
\centering
\includegraphics[width=\columnwidth]{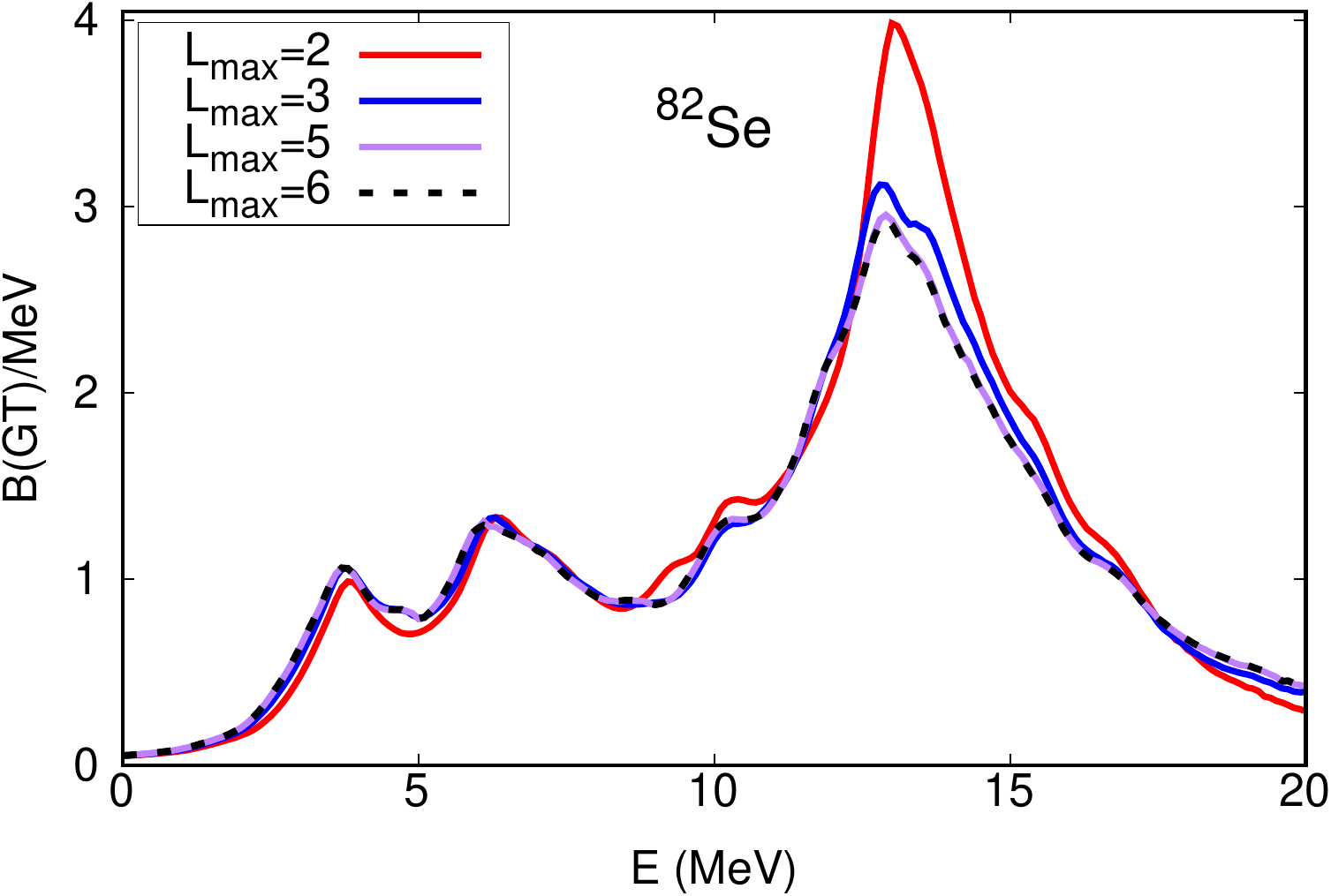}
\caption{Gamow-Teller strength distribution in $^{82}$Se obtained in the pnFAM*,
with several values for the maximum angular momentum $L_\text{max}$ in states
excited by the like-particle operators in Eq.\ \eqref{eq:Ghere}.  No other
cutoffs are imposed, i.e.\ $v_\text{min} = 0$.}
\label{fig:se-L}
\end{figure}

To conclude this section, we look at the effects of limiting ourselves to
like-particle excitation operators with $L \leq 6$ in Eq.\ \eqref{eq:Ghere}.
Figure \ref{fig:se-L} shows strength distributions in $^{82}$Se for different
values $L_\text{max}$ of the maximum angular momentum in the excitation
operators.  One can see that by $L_\text{max}=5$ the distribution has converged.
The pattern is almost exactly the same in other two isotopes we examined
earlier.

\section{$\beta$-decay rates}

As we've noted, $\beta$-decay rates depend sensitively on details of the
low-lying parts of strength distributions.  Unlike gross features such as giant
resonances, low-lying strength is strongly affected by the isoscalar pairing
interaction, the coefficient of which was fit in global Skyrme-QRPA calculations
of $\beta$ decay \cite{Mustonen2016,Ney20}.  As we have seen, allowing
quasiparticles to couple to phonons also alters low-lying strength.  So that we
do not confuse the effects, we continue to work with SGII, without isoscalar
pairing. 

\begin{table}[b]
\caption{Deformation parameter, experimental half-life in seconds, pnFAM 
half-life with the Skyrme functional SGII, pnFAM* half-life with the same functional,
and pnFAM* half-life when phonons with $v_N < 0.3$ are excluded, for 11 deformed
isotopes.} 
\begin{ruledtabular}
   \begin{tabular}{cccccc}
   Isotope & $\beta$ & $t_{1/2}^{\text{Exp.}}$(s) & $t_{1/2}^\text{pnFAM}$(s)
   & $t_{1/2}^{\text{pnFAM*}}$(s) & $t_{1/2}'^{\text{\ pnFAM*}}$(s)\\[.05in]
   \hline &&&&&\\[-.1in]
   $^{78}$Zn & 0.12 & 1.47 & 408 & 3.77 & 4.80\\
   $^{168}$Gd & 0.31 & 3.03 & 381 & 37.1 & 39.7\\ 
   $^{152}$Ce & 0.29 & 1.40 & 93.1 & 19.0& 20.3 \\
   $^{156}$Nd & 0.32 & 5.49 & 470 & 53.5 & 59.2\\
   $^{164}$Sm & 0.33 & 1.42 & 142 & 17.2 & 18.5\\
   $^{154}$Ce & 0.30 & 0.30 & 19.2 & 7.26& 7.89\\
   $^{112}$Mo & $-0.18$ & 0.15 & 1.92 & 2.47& 2.31 \\
   $^{94}$Kr & $-0.22$ & 0.21 & 1.48 & 3.23& 3.01 \\
   $^{112}$Ru & $-0.21$ & 1.75 & 93 & 27.0& 31.0\\
   $^{106}$Mo & $-0.20$ & 8.73 & 62.8 & 38.0& 49.6\\
   $^{96}$Sr & $-0.21$ & 1.07 & 23.8 & 20.0& 25.9\\
   \end{tabular}
\end{ruledtabular}
\label{tab:rates}
\end{table}

Table \ref{tab:rates} shows the effects of the quasiparticle-phonon coupling on
decay rates in 11 representative heavy nuclei.  Six of the 11 are prolate and
five are oblate (though data in $^{106}$Mo, which we find to be oblate, may be
more consistent with prolate deformation \cite{Ha20}).  In all the isotopes, the
pnFAM without any modification over-predicts the lifetimes by placing too little
strength at low energies.  In the prolate cases, the quasiparticle-phonon
coupling always reduces the lifetime, often dramatically, and always improves
the agreement with experiment.  Though the effect is smaller, the agreement is
usually better in the oblate isotopes as well.  In two of them, however, the
coupling actually increases the lifetime slightly, making the agreement with
experiment a little worse.  Because most of the allowed rates increase in the
pnFAM*, the forbidden contributions to the decay, which increase less, are a
smaller percentage of the total rate than in the pnFAM itself. 

In prior global QRPA calculations, isoscalar pairing was used to increase the
strength below the $\beta$-decay $Q$ value, decreasing predicted lifetimes so
that on average they were correct.    Many were significantly too small,
however, and many others significantly too large, in spite of the isoscalar
pairing.  With the coupling of quasiparticles to phonons also increasing
lifetimes (usually), in a way that is more natural, isoscalar pairing can be
weakened.  Though we don't know for sure that the fluctuations in computed rates
will be smaller than before, work in spherical nuclei \cite{Niu2018} suggests
that they will.

\section{Discussion}

Our method for including the coupling of quasiparticles to phonons is much more
efficient than the extension of the matrix QRPA.  It is not yet efficient
enough, however, to allow us to apply it to all of the approximately 4000
neutron-rich isotopes.   At present, we extract the quasiparticle-phonon
coupling by first computing like-particle strength functions at many values of
$\omega$ and in many channels, identifying peaks, and employing Eqs.\
\eqref{eq:vertex} and \eqref{eq:GN0} for each one.  How might we get the same
information more efficiently?

One advantage of the traditional matrix QRPA is that it directly produces the
transition amplitudes $\mathcal{X}^N_{\alpha \beta}$ and $\mathcal{Y}^N_{\alpha
\beta}$ [see Eq.\ \eqref{eq:extractxy}], which in turn determine
$\braket{\beta|H|\beta_1,N}$ and therefore $W_{\pi \nu,\pi' \nu'}(\omega)$ in a
straightforward algebraic way.  The chief disadvantage is the long time it takes
to construct the many elements of the QRPA Hamiltonian matrix.  The FAM can
actually be used to construct the matrix much faster \cite{Avogadro13}, and a
starting point for future work is to build and diagonalize the matrix, use the
results and Eq.\ \eqref{eq:VN} to select the most important phonons, and then
construct $W$.   And we can do even better by employing a Lanczos-like approach
to produce only the most important parts of the matrix.  Ref.\ \cite{Toivanen10}
applies the Arnoldi algorithm in concert with a FAM-like procedure to obtain
matrices with dimensions of order $100 \times 100$ that, when diagonalized,
reproduce strength distributions accurately.   Another promising option is to
use expand the QRPA $\mathcal{X}$'s and $\mathcal{Y}$'s in terms of FAM
amplitudes at values of $\omega$ far from the real axis.  This approach, which
we will present in a separate paper \cite{Hinohara23}, is analogous to the
eigenvalue continuation introduced in Ref.\ \cite{Frame18}.

Once we have a faster procedure, we will be able use the pnFAM* to refit the
parameters of the time-odd part of any Skyrme functional to a set of selected
$\beta$-decay rates, charge-changing resonance energies, etc., as was done with
the pnFAM itself in Ref.\ \cite{Mustonen2016}.  We can also include in the fit
the parameters multiplying chiral two-body weak-current operators, the
infrastructure for which was presented in Ref.\ \cite{Ney22}.  The result should
be much better predictions for $r$-process simulations.

This plan raises the question of what it means to extend density-functional
theory beyond the QRPA, i.e.\ beyond a time-dependent mean-field ansatz.  Does
it make sense to use a density-dependent ``Hamiltonian'' in conjunction with
beyond-mean-field correlations?  Aren't the correlations already implicit in the
functional itself?  We need not answer these questions carefully to be confident
that including more correlations is worthwhile.  Doing so pushes our method in
the direction of an ab initio solution of the Schr\"odinger equation.  Double
counting of correlations can be removed through parameter refitting.  The more
correct physics we include by extending our many-body methods, the less we
burden the functional with mocking that physics up.  An exact many-body method
would be appropriate in conjunction with a functional that corresponds simply to
the mean-field expectation value of an \textit{ab initio} Hamiltonian.  Our long
term goal is to push nuclear density-functional theory as close to that point as
possible.  The improved description of charge-changing processes presented here
is a good start.\\

\section*{\label{sec:acknowledgments}Acknowledgments}

Many thanks to Elena Litvinova and Yinu Zhang for helpful discussions.  This
work was supported in part by the US Department of Energy under Contracts
DE-FG02-97ER41019 and DE-SC0023175 (NUCLEI SciDAC-5 collaboration).  The work of
MK was partly supported by the Academy of Finland under the Academy Project No.
339243.  The work of NH was supported by the JSPS KAKENHI (Grants No.
JP19KK0343, No. JP20K03964, and No. JP22H04569).


%

\end{document}